\documentclass{article}
\title{Conditions for 3-partite and 4-partite genuine entanglement}
\author{Mark Hillery \\ Department of Physics and Astronomy \\ Hunter College of CUNY \\ 695 Park Avenue \\New York, NY 10065 \\ and \\ Physics Program \\ Graduate Center of CUNY \\ 365 Fifth Avenue \\ New York, NY 10016}

\begin{document}
\maketitle

\begin{abstract}
A system of three or four particle can be entangled in a number of different ways.  It may be the case that only subsets of the particles are entangled, and these subsets are not entangled with each other.  It may also be the case that the state is the sum of states in which entanglement only exists within subsets.  If this is not the case, the state is said to be genuinely entangled.  GHZ states, for example, are genuinely entangled.  Deciding whether a state is genuinely entangled is not simple, but conditions do exist to detect it.  Here we would like to propose additional sufficient conditions based on one for bipartite entanglement.
\end{abstract}

ORCID ID: 0000-0002-1296-075X

\section{Introduction}
As is to be expected, multipartite entanglement is considerably more complicated than bipartite entanglement.  For example, a multipartite state may consist of subsets that are entangled within themselves but are not entangled with each other.  A state with entanglement that is more multipartite in character would be one that is not separable across any bipartite partition.  An even more stringent requirement is that the state, in this case the density matrix, is not the sum of states that are separable across a bipartite partition, but with each of the states being separable across a different bipartite partition.  Such states are called genuinely multipartite entangled.

The first condition for detecting genuine tripartite nonlocality is due to Svetlichny \cite{svetlichny1}.  He found an inequality that when violated by a state guarantees that the state is genuinely tripartite nonlocal.  This was subsequently extended to general multipartite states and entanglement; the inequalities could detect both genuine multipartite nonlocality and genuine multipartite entanglement \cite{svetlichny2}.  It should be noted that in the tripartite case, both GHZ and W states are genuinely tripartite entangled. Since this initial work, a number of criteria for detecting genuine entanglement have been developed \cite{huber1}-\cite{Li}.  We would like to show how simple additional conditions can be derived based on a bipartite entanglement condition derived in \cite{hillery}.

We will begin by considering the tripartite case.  We have 3 systems, $a$, $b$, and $c$.  A quantum state of these systems is genuinely tripartite entangled if its density matrix cannot be written in the form
\begin{equation}
\label{tripartite}
\rho_{abc}= p_{1}\rho^{(1)}_{ab|c} + p_{2}\rho^{(2)}_{ac|b} + p_{3} \rho^{(3)}_{bc|a} ,
\end{equation}
where $p_{j}\geq 0$ for $j=1,2,3$, $p_{1}+p_{2}+p_{3}=1$, and $\rho^{(1)}_{ab|c}$ is a density matrix that is separable across the $ab$-$c$ partition, that is, in $\rho^{(1)}_{ab|c}$ the systems $ab$ and $c$ are not entangled.  The density matrices $\rho^{(2)}_{ac|b}$ and $\rho^{(3)}_{bc|a}$ are defined similarly.  We want to find a sufficient condition for a density matrix to be genuinely tripartite entangled, and we will do this by finding a condition that $\rho_{abc}$ in the above equation must satisfy, and then any state that violates that condition must be genuinely tripartite entangled. 

In \cite{hillery} two conditions for separability in a bipartite system were proved.  If $\rho_{ab}$ is a separable bipartite density matrix for systems $a$ and $b$, then the following conditions hold
\begin{eqnarray}
\label{inequalities}
|\langle L_{a}^{\dagger}M_{b}\rangle |^{2} & \leq & \langle L_{a}^{\dagger}L_{a}M_{b}^{\dagger}M_{b}\rangle \nonumber  \\
|\langle L_{a}M_{b}\rangle |^{2} & \leq & \langle L_{a}^{\dagger}L_{a}\rangle \langle M_{b}^{\dagger}M_{b}\rangle ,
\end{eqnarray}
where $L_{a}$ and $M_{b}$ are arbitrary non-hermitian operators on systems $a$ and $b$ respectively (the conditions hold for hermitian operators too, but are useless in that case because they then hold for all density matrices).  In \cite{agusti} the second of these conditions was used to show that if a tripartite density matrix is of the form given in Eq.\ (\ref{tripartite}), then it must obey
\begin{equation}
\label{cond1}
|\langle ABC\rangle |\leq \sqrt{\langle A^{\dagger}A\rangle\langle B^{\dagger}BC^{\dagger}C\rangle} + \sqrt{\langle B^{\dagger}B\rangle\langle A^{\dagger}AC^{\dagger}C\rangle} + \sqrt{\langle C^{\dagger}C\rangle\langle A^{\dagger}B^{\dagger}B\rangle} 
\end{equation}
where $A$, $B$, and $C$ are operators on systems $a$, $b$, and $c$ respectively.  If this condition is violated, the state is genuinely tripartite entangled.  This was proved assuming $A$, $B$, and $C$ are mode annihilation operators, but the proof does not depend on this assumption.  They could be, for example, spin lowering operators. Now let's use the first condition in Eq.\ (\ref{inequalities}) to derive another condition for genuinely tripartite entanglement.

We will consider $\langle A^{\dagger}BC\rangle = {\rm Tr}(A^{\dagger}BC\rho )$, where $\rho$ is of the form given in Eq.\ (\ref{tripartite}). Now because the density matrix $\rho^{(1)}_{ab|c}$ is separable across the $ab$-$c$ partition, we can apply the first condition in Eq.\ (\ref{inequalities}) with $L^{\dagger}=A^{\dagger}B$ and $M=C$ to give
\begin{equation}
\label{ineq1}
| \langle A^{\dagger}BC\rangle_{1}| = | {\rm Tr}( A^{\dagger}BC \rho^{(1)}_{ab|c} )| \leq  \langle A^{\dagger}ABB^{\dagger} C^{\dagger}C\rangle_{1}^{1/2} .
\end{equation}
Now note that because $A^{\dagger}ABB^{\dagger} C^{\dagger}C$ is a positive operator,
\begin{equation}
\label{ineq2}
p_{1} {\rm Tr}( \rho^{(1)}_{ab|c}  A^{\dagger}ABB^{\dagger} C^{\dagger}C) \leq {\rm Tr}(\rho A^{\dagger}ABB^{\dagger} C^{\dagger}C) .
\end{equation}
The same procedure can be applied to the remaining two terms.  For $\rho^{(2)}_{ac|b}$ we choose $L^{\dagger} = A^{\dagger}C$ and $M=B$ yielding
\begin{equation}
p_{2} | \langle A^{\dagger}BC\rangle_{2} |\leq [ {\rm Tr}(\rho A^{\dagger}A B^{\dagger}B CC^{\dagger}) ]^{1/2} .
\end{equation}
For the $ \rho^{(3)}_{bc|a}$ term we choose $L^{\dagger}= A^{\dagger}$ and $M=BC$ giving 
\begin{equation}
p_{3} | \langle A^{\dagger}BC\rangle_{3} |\leq [ {\rm Tr}(\rho A^{\dagger}AB^{\dagger}BC^{\dagger}C) ]^{1/2} .
\end{equation}
Putting this all together, we find that if the tripartite density matrix is of the form given in Eq.\ (\ref{tripartite}), then
\begin{eqnarray}
\label{cond2}
| \langle  A^{\dagger}BC\rangle | & \leq & \langle A^{\dagger}ABB^{\dagger} C^{\dagger}C\rangle^{1/2} + \langle A^{\dagger}A B^{\dagger}B CC^{\dagger} \rangle^{1/2} \nonumber \\
&&+ \langle A^{\dagger}AB^{\dagger}BC^{\dagger}C \rangle^{1/2} .
\end{eqnarray}
If this condition is violated the state is genuinely tripartite entangled.  

We can actually do a bit better.  From Eq.\ (\ref{ineq1}), and the fact that the operator is positive, instead of Eq.\ (\ref{ineq2}) we have
\begin{equation}
p_{1} {\rm Tr}( \rho^{(1)}_{ab|c}  A^{\dagger}ABB^{\dagger} C^{\dagger}C) \leq p_{1} {\rm Tr}(\rho A^{\dagger}ABB^{\dagger} C^{\dagger}C) .
\end{equation}
Similar relations hold for the other two terms.  This then implies that
\begin{eqnarray}
\label{stronger}
| \langle  A^{\dagger}BC\rangle | & \leq & p_{1} \langle A^{\dagger}ABB^{\dagger} C^{\dagger}C\rangle^{1/2} +  p_{2} \langle A^{\dagger}A B^{\dagger}B CC^{\dagger} \rangle^{1/2} \nonumber \\
&&+ p_{3} \langle A^{\dagger}AB^{\dagger}BC^{\dagger}C \rangle^{1/2}  \nonumber \\
& \leq & \max ( \langle A^{\dagger}ABB^{\dagger} C^{\dagger}C\rangle^{1/2},\,  \langle A^{\dagger}A B^{\dagger}B CC^{\dagger} \rangle^{1/2}, \nonumber \\ 
&&  \langle A^{\dagger}AB^{\dagger}BC^{\dagger}C \rangle^{1/2} ) .  
\end{eqnarray}
Since the condition in Eq.\ (\ref{stronger}) is clearly stronger than that in Eq.\ (\ref{cond2}), we will be focussing on it. We should note that the same reasoning applies to the inequality in Eq.\ (\ref{cond1}), which then becomes
\begin{eqnarray}
\label{stronger2}
|\langle ABC\rangle | & \leq & \max \left(\sqrt{\langle A^{\dagger}A\rangle\langle B^{\dagger}BC^{\dagger}C\rangle} , \sqrt{\langle B^{\dagger}B\rangle\langle A^{\dagger}AC^{\dagger}C\rangle}, \right. \nonumber \\
&& \left. \sqrt{\langle C^{\dagger}C\rangle\langle A^{\dagger}B^{\dagger}B\rangle} \right) .
\end{eqnarray}
This was shown for mode creation and annihilation operators in \cite{jarvis}, but, again, the proof does not depend on this choice.

Let's look at two simple examples.  If we apply the condition in Eq.\ (\ref{stronger2}) to the state $|\psi_{1}\rangle = c_{0} |000\rangle + c_{1}|111\rangle$, with $A$, $B$, and $C$ being $\sigma^{(-)}$ operators, we find that the condition is violated if $|c_{0}| > |c_{1}|$. If we choose $|\psi_{2}\rangle = c_{0}|011\rangle + c_{1}|100\rangle$ with the same choice of $A$, $B$, and $C$ and apply the condition in Eq.\ (\ref{cond2}) or Eq.\ (\ref{stronger}), then we find the condition is violated for any nonzero choice of $c_{0}$ and $c_{1}$.  This suggests the condition in Eq.\ (\ref{stronger}) is stronger than the condition in Eq.\ (\ref{stronger2}).

\section{Examples for tripartite systems}
Let's look at qubits some more, and then move on to qutrits.  Setting $A=|0\rangle\langle 1|$, $B=|1\rangle\langle 0|$, and $C=|0\rangle\langle 1|$, we find $A^{\dagger}BC=|110\rangle\langle 001|$.  We also find that
\begin{eqnarray}
A^{\dagger}ABB^{\dagger}C^{\dagger}C & = & |111\rangle \langle 111| ,\nonumber \\
A^{\dagger}AB^{\dagger}BCC^{\dagger} & = & |100\rangle\langle 100| \nonumber \\
A^{\dagger}AB^{\dagger}BC^{\dagger}C & = & |101\rangle \langle 101| 
\end{eqnarray}
Note that the states $|000\rangle$, $|001\rangle$, $|010\rangle$, $|011\rangle$, and $|110\rangle$ are all annihilated by the three operators in the above equation.  That means that any linear combination of them will give zero when substituted into the right hand side of our inequality, Eq.\ (\ref{stronger}).  The only two of these states that are connected by the operator $A^{\dagger}BC$ are $|001\rangle$ and $|110\rangle$.  Therefore, if we substitute the state
\begin{equation}
c_{1}|000\rangle +c_{2}|001\rangle +c_{3}|010\rangle + c_{4}|011\rangle + c_{5}|110\rangle 
\end{equation}
into our inequality the left-hand side will be $|c_{2}c_{5}|$ and the right-hand side will be zero.  Therefore, all states of this type are genuinely tripartite entangled.

Another possibility is to consider a state of the form
\begin{equation}
|\psi\rangle = c_{0}|0uu\rangle + c_{1}|u00\rangle ,
\end{equation}
where $|u\rangle$ is a state such that $\langle 1|u\rangle =s$, and $s$ is real and positive.  Again choosing $A=B=C=\sigma^{(-)}$, we find that the expectation values on the right-hand side of Eq.\ (\ref{stronger}) are zero, and the left-hand side is $|c_{1}^{\ast}c_{0}s^{3}|$, so that this state is also genuinely tripartite entangled.

It is also useful to look at mixed states.  Suppose that $A=B=C=\sigma^{(-)}$, $|\psi\rangle = c_{0}|011\rangle + c_{1}|100\rangle$, and
\begin{equation}
\rho = s|\psi\rangle\langle\psi | + \frac{1-s}{8} I ,
\end{equation}
where $0\leq s \leq 1$.  The expectation values on the right-hand side of Eq.\ (\ref{stronger}) are all equal to $(1-s)/8$, and we then have that the state is genuinely entangled if
\begin{equation}
s ^{2}|c_{0}^{\ast}c_{1}|^{2} > \frac{1-s}{8} .
\end{equation}
In the case that $c_{0}=c_{1}=1\sqrt{2}$ this becomes $s > 1/2$.

Now let's move on to qutrits.  Our qutrit basis is $\{ |0\rangle , |1\rangle , |2\rangle \}$, and for our operators $A$, $B$, and $C$ we will choose
\begin{eqnarray}
A & = & |0\rangle_{a}\langle 1| + |1\rangle_{a}\langle 2| \nonumber \\
B & = & |1\rangle_{b}\langle 0| + |2\rangle_{b}\langle 1| \nonumber \\
C & = & |0\rangle_{c}\langle 1| + |1\rangle_{c}\langle 2| .
\end{eqnarray}
Note that $A$ and $C$ are lowering operators, and $B$ is a raising operator.  The operator $A^{\dagger}BC$ is nonzero only on the $8$ basis states of the form $(|0\rangle\ {\rm or}\ |1\rangle )\otimes (|0\rangle\ {\rm or}\ |1\rangle ) \otimes (|1\rangle\ {\rm or}\ |2\rangle )$.  It transforms these states as follows
\begin{eqnarray}
|001\rangle \rightarrow |110\rangle & |011\rangle \rightarrow |120\rangle & |101\rangle \rightarrow |210\rangle \nonumber \\
|002\rangle \rightarrow |111\rangle & |012\rangle \rightarrow |121\rangle & |102\rangle \rightarrow |211\rangle \nonumber \\
|111\rangle \rightarrow |220\rangle & |112\rangle \rightarrow |221\rangle & 
\end{eqnarray}
The operators on the right-hand side of Eq.\ (\ref{stronger}) are
\begin{eqnarray}
\label{3trit}
A^{\dagger}ABB^{\dagger}C^{\dagger}C & = & (|1\rangle\langle 1|+|2\rangle\langle 2|) \otimes (|1\rangle\langle 1|+|2\rangle\langle 2|) \otimes (|1\rangle\langle 1|+|2\rangle\langle 2|) \nonumber \\
A^{\dagger}AB^{\dagger}BCC^{\dagger} & = & (|1\rangle\langle 1|+|2\rangle\langle 2|) \otimes (|0\rangle\langle 0|+|1\rangle\langle 1|) \otimes (|0\rangle\langle 0|+|1\rangle\langle 1|) \nonumber \\
A^{\dagger}AB^{\dagger}BC^{\dagger}C & = & (|1\rangle\langle 1|+|2\rangle\langle 2|) \otimes (|0\rangle\langle 0|+|1\rangle\langle 1|) \otimes (|1\rangle\langle 1|+|2\rangle\langle 2|) . \nonumber \\
\end{eqnarray}
The states that are annihilated by all three of these operators are of the form $|0jk\rangle$, where $j,k=0,1,2$ and the states $|120\rangle$ and $|220\rangle$.  It is also the case that states of the form $|12j\rangle$, where $j=0,1,2$, are annihilated by the second and third operators.  Taking advantage of these properties, we see that if we put the state $c_{0}|011\rangle + c_{1}|120\rangle$ into Eq.\ (\ref{stronger}), we find that the state is genuinely if $|c_{1}^{\ast}c_{0}| >0$.  We can add any of the states that are annihilated by all three operators to this state, and the condition will not change.

Let's try two possibilities that involve some states that are not all annihilated by all three operators.  Consider a state of the form
\begin{equation}
|\psi\rangle = c_{0}|002\rangle + c_{1}|111\rangle + c_{2}|220\rangle .
\end{equation}
The first and third state are annihilated by the operators in Eq.\ (\ref{3trit}), and the middle state is annihilated by none of them.  All three of these states are connected by the operator $A^{\dagger}BC$, which maps $|002\rangle \rightarrow |111\rangle \rightarrow |220\rangle$.  Substituting this state into our inequality, we find that the state is genuinely tripartite entangled if $|c_{1}^{\ast}c_{0} + c_{2}^{\ast}c_{1}| > |c_{1}|$, which is possible to satisfy.  For example, suppose $c_{0}$, $c_{1}$, and $c_{2}$ are real and positive, and $c_{1}\neq 0$.  Then the condition is satisfied if $c_{0} + c_{2} > 1$.  If $c_{0}$ and $c_{2}$ are greater than $1/2$ and less than $1/\sqrt{2}$ this will be satisfied, and we will have $1/2 > c_{1}^{2} > 0$.


\section{Down conversion}

As an application of these ideas, let us turn to an example from nonlinear optics.  In nonlinear optics, the process of parametric down conversion entails a photon at frequency $\omega_{1} = \omega_{2}+\omega_{3}$ splitting into two photons, one at $\omega_{2}$ and one at $\omega_{3}$.  The reverse process is also allowed, a photon at $\omega_{2}$ and a photon at $\omega_{3}$ combining into a photon at $\omega_{1}$.  The Hamiltonian governing this process is
\begin{equation}
H=\omega_{1} a^{\dagger}a + \omega_{2}b^{\dagger}b + \omega_{3} c^{\dagger}c + g (a^{\dagger}bc + ab^{\dagger}c^{\dagger}) .
\end{equation}
Here $a$, $b$, and $c$ are the annihilation operators for the modes at $\omega_{1}$, $\omega_{2}$, and $\omega_{3}$, respectively.  The coupling constant in the interaction term, $g$ is proportional to the nonlinear susceptibility, $\chi^{(2)}$, of the material in which the interaction takes place.  This Hamiltonian obeys a conservation law, the operator $2N_{a}+N_{b}+N_{c}$ commutes with the Hamiltonian, where $N_{a}$, $N_{b}$, and $N_{c}$ are the number operators for their respective modes.  A basis for the Hilbert space for this system is $\{ |n_{a},n_{b},n_{c}\rangle\, |\, n_{a},n_{b}, n_{c} = 0,1,\ldots \}$, where the numbers $n_{a}$, $n_{b}$, and $n_{c}$ are the numbers of photons in each of the modes.  If we start the system in the state $|N,0,0\rangle$, due to the conservation law at a later time the state will be of the form
\begin{equation}
|\psi\rangle = \sum_{n=0}^{N} c_{n} |N-n, n, n\rangle .
\end{equation}
We want to use our conditions to investigate whether the state $|\psi\rangle$ is genuinely tripartite entangled.

Let us divide the basis states in $|\psi\rangle$ into disjoint sets $\{ |N,0,0\rangle ,\, |N-1,1,1\rangle\}$, $\{ |N-2,2,2\rangle , \, |N-3,3,3\rangle \}$, \ldots $\{ |2, N-2,N-2\rangle ,\, |1, N-1, N-1\rangle \}$ and assume for convenience that $N$ is even.  For the set $\{ |N-n,n,n\rangle ,\, |N-n-1,n+1,n+1\rangle\}$, $n$ even, define $A_{n}=|N-n-1\rangle_{a}\langle N-n|$, $B_{n}=|n\rangle_{b}\langle n+1|$, and $C=|n\rangle_{c}\langle n+1|$.  The operator $A^{\dagger}_{n}B_{n}C_{n} = |N-n,n,n\rangle\langle N-n-1,n+1,n+1 |$ has the expectation value $c_{n}^{\ast}c_{n+1}$ in the state $|\psi\rangle$.  The operators
\begin{eqnarray}
A_{n}^{\dagger}A_{n}B_{n}B_{n}^{\dagger}C_{n}^{\dagger}C_{n} & = & |N-n, n, n+1\rangle \langle N-n,n,n+1| \nonumber \\
A_{n}^{\dagger}A_{n}B_{n}^{\dagger}B_{n}C_{n}C_{n}^{\dagger} & = & |N-n,n+1,n\rangle\langle N-n,n+1, n| \nonumber \\
A_{n}^{\dagger}A_{n}B_{n}^{\dagger}B_{n}C_{n}^{\dagger}C_{n} & = & |N-n,n+1,n+1\rangle\langle N-n,n+1,n+1| .
\end{eqnarray}
all have the property that their expectation value in the state $|\psi\rangle$ is zero.  Therefore, if any of the of the products $|c_{n}^{\ast}c_{n+1}|$, for $n$ even, is greater than zero, the the state is genuinely tripartite entangled.  If we wish, we can define $A=\sum_{n=0,\, n\, {\rm even}}^{N-2} A_{n}$ and similarly for $B$ and $C$.  Our condition then tells us that if
\begin{equation}
\left| \sum_{n=0,\, n\, {\rm even}}^{N-2} c_{n}^{\ast}c_{n+1}\right| > 0 ,
\end{equation}
then the state is genuinely tripartite entangled.

\section{Genuine 4-partite entanglement}
We can extend the reasoning in the previous sections to larger systems.  Here we will do so for a 4-partite system and eventually look at the case in which the subsystems are qubits.  We will denote the qubits by $a$, $b$, $c$, and $d$, with operator $A$ acting on qubit $a$, and similarly for the other qubits.  A state is genuinely 4-partite entangled if it is not of the form
\begin{eqnarray}
\label{4-partite}
\rho & = & p_{1}\rho^{(1)}_{a|bcd} + p_{2}\rho^{(2)}_{b|acd} + p_{3}\rho^{(3)}_{c|abd} + p_{4}\rho^{(4)}_{d|abc} + p_{5}\rho^{(5)}_{ab|cd} \nonumber \\
&& + p_{6}\rho^{(6)}_{ac|bd} + p_{7}\rho^{(7)}_{ad|bc} .
\end{eqnarray}
The terms in the above equation represent all possible bi-partitions of the four qubits, and $\sum_{j=1}^{7} p_{j} = 1$.  We only need to rule out partitions into two subsets, because this rules out partitions into more subsets as well.
 
We will consider an operator of the form $A^{\dagger}BCD$ and apply the first inequality in Eq.\ (\ref{inequalities}).  We will only work out some of the terms, but the others follow in a very similar fashion.  Setting $L_{a}^{\dagger}=A^{\dagger}CD$ and $M_{b}=B$, we find that for $\rho^{(2)}_{b|acd}$ 
\begin{equation}
|\langle A^{\dagger}BCD\rangle_{2}| \leq \langle A^{\dagger}AB^{\dagger}BCC^{\dagger}DD^{\dagger}\rangle_{2}^{1/2} \leq  \langle A^{\dagger}AB^{\dagger}BCC^{\dagger}DD^{\dagger}\rangle^{1/2} ,
\end{equation}
where, as before, the subscript on the brackets indicates that the expectation value is with respect to $\rho^{(2)}_{b|acd}$ and brackets with no subscript indicate that the expectation value is with respect to $\rho$.  If we now look at $\rho^{(5)}_{ab|cd}$ with $L_{a}^{\dagger}=A^{\dagger}B$ and $M_{b}=CD$, we find
\begin{equation}
|\langle A^{\dagger}BCD\rangle_{5}| \leq \langle A^{\dagger}ABB^{\dagger}C^{\dagger}CD^{\dagger}D\rangle_{5}^{1/2} \leq  \langle A^{\dagger}ABB^{\dagger}C^{\dagger}CD^{\dagger}D\rangle^{1/2} .
\end{equation}
Proceeding as before, we find that if $\rho$ is of the form given in Eq.\ (\ref{4-partite}), then $|\langle A^{\dagger}BCD\rangle|$ must be less than or equal to the maximum of the seven quantities
\begin{eqnarray}
 \langle A^{\dagger}AB^{\dagger}BC^{\dagger}CD^{\dagger}D\rangle^{1/2} , &  \langle A^{\dagger}AB^{\dagger}BCC^{\dagger}DD^{\dagger}\rangle^{1/2} , &  \langle A^{\dagger}ABB^{\dagger}C^{\dagger}CDD^{\dagger}\rangle^{1/2} , \nonumber \\
  \langle A^{\dagger}ABB^{\dagger}CC^{\dagger}D^{\dagger}D\rangle^{1/2} , &  \langle A^{\dagger}ABB^{\dagger}C^{\dagger}CD^{\dagger}D\rangle^{1/2} , &  \langle A^{\dagger}AB^{\dagger}BCC^{\dagger}D^{\dagger}D\rangle^{1/2} , \nonumber \\ 
 \langle A^{\dagger}AB^{\dagger}BC^{\dagger}CDD^{\dagger}\rangle^{1/2} . & & 
  \end{eqnarray}
  
  To get an idea of how this works, let's look at the case or four qubits in which the operators are $A=B=C=D=\sigma^{(-)}$.  Then the operator $A^{\dagger}BCD$ couples the states $|0111\rangle$ and $|1000\rangle$, and each of the seven operators appearing in the previous equation is a projection onto a single state.  The corresponding states are
  \begin{eqnarray}
  |1111\rangle & |1100\rangle & |1010\rangle \nonumber \\
  |1001\rangle & |1011\rangle & |1101\rangle \nonumber \\
  |1110\rangle .& & 
  \end{eqnarray}
  That means that any linear combination of the nine basis states orthogonal to the seven above, in which the states $|0111\rangle$ and $|1000\rangle$ are both present, will be genuinely 4-partite entangled.

As before, we can look at a mixed state, in particular
\begin{equation}
\rho = s|\psi\rangle\langle\psi | + \frac{1-s}{16} I ,
\end{equation}
where now $|\psi\rangle = c_{0}|0111\rangle + c_{1}|1000\rangle$, and, as before, $0\leq s \leq 1$.  The condition for genuine 4-partite entanglement for this state becomes
\begin{equation}
s ^{2}|c_{0}^{\ast}c_{1}|^{2} > \frac{1-s}{16} .
\end{equation}
In the case $c_{0}=c_{1}=1/\sqrt{2}$ this gives $s > (\sqrt{17}-1)/8 = 0.39$, which is a larger range of $s$ than in the tripartite case.

\section{Conclusion}
Using an entanglement condition derived in \cite{hillery} we have derived conditions for genuine 3- and 4-partite entanglement.  These were applied to qubit and qutrit states, and an example from nonlinear optics was discussed.

\section*{Acknowledgments}
This research was supported by NSF grant Collaborative Research: NeTS: Medium 2504622.

\end{document}